\documentclass[lettersize,journal]{IEEEtran}
\usepackage{amsmath,amsfonts,amsthm,amssymb,mathrsfs}
\usepackage{algorithm}
\usepackage{algpseudocode}
\usepackage{array}
\usepackage[caption=false,font=scriptsize]{subfig}
\usepackage{textcomp}
\usepackage{stfloats}
\usepackage{url}
\usepackage{verbatim}
\usepackage{graphicx}
\usepackage{bbding}
\usepackage{multirow}
\usepackage{cite, bm, bbm, xcolor}
\usepackage{refcount}
\usepackage{flafter}
\usepackage{makecell}
\usepackage{diagbox}
\usepackage{booktabs}
\usepackage{graphicx}

\newtheoremstyle{myremark}
  {2pt}
  {2pt}
  {\normalfont}
  {}
  {\bfseries}
  {}
  {1em}
  {\thmname{#1}~\thmnumber{#2}\thmnote{ (#3)}} 
\theoremstyle{myremark}
\newtheorem{remark}{Remark}

\newtheorem{lemma}{Lemma}
\theoremstyle{definition}

\usepackage{ulem}
\usepackage[T1]{fontenc}
\providecommand{\url}[1]{#1}
\allowdisplaybreaks[4]

\title{User Localization and Channel Estimation for Pinching-Antenna Systems (PASS)}
\author{
Xiaoxia Xu, \textit{Member, IEEE}, 
Xidong Mu,  \textit{Member, IEEE}, 
Yuanwei Liu, \textit{Fellow, IEEE}, 
Hong Xing,   \textit{Member, IEEE}, 
and Arumugam Nallanathan, \textit{Fellow, IEEE}
\vspace{-1em}
\thanks{X. Xu and A. Nallanathan are with the School of Electronic Engineering and Computer Science, Queen Mary University of
London, London E1 4NS, U.K. (email: \{x.xiaoxia, a.nallanathan\}@qmul.ac.uk).}
\thanks{X. Mu is with the Centre for Wireless Innovation (CWI), Queen's University Belfast, Belfast, BT3 9DT, U.K. (x.mu@qub.ac.uk).}
\thanks{Y. Liu is with the Department of Electrical and Electronic Engineering (EEE), The University of Hong Kong, Hong Kong (e-mail: yuanwei@hku.hk).}
\thanks{H. Xing is with the IoT Thrust, The Hong Kong University of Science and Technology (Guangzhou), Guangzhou, 511453,
China, and also affiliated with the Department of ECE, The Hong Kong University of Science and Technology, HK SAR (e-mail: hongxing@ust.hk).}
}

\date{\today}

\begin{document}

\maketitle

\begin{abstract}
    This letter proposes a novel user localization and channel estimation framework for pinching-antenna systems (PASS),  
    where pinching antennas are grouped into subarrays on each waveguide to cooperatively estimate user/scatterer locations, thus reconstructing channels.
    Both single-waveguide (SW) and multi-waveguide (MW) structures are considered. 
    SW consists of multiple alternatingly activated subarrays, while MW deploys one subarray on each waveguide to enable concurrent subarray measurements. 
    For the 2D scenarios with a fixed user/scatter height, an orthogonal matching pursuit-based geometry-consistent localization (OMP-GCL) algorithm is proposed, 
    which leverages inter-subarray geometric relationships and compressed sensing for precise estimation. 
    Theoretical analysis on Cram\'{e}r-Rao lower bound (CRLB) demonstrates that: 
    1) The estimation accuracy can be improved by increasing the geometric diversity through multi-subarray deployment; and 
    2) SW provides a limited geometric diversity within a $180^{\circ}$ half space and leads to angle ambiguity, while MW enables full-space observations and reduces overheads. 
    The OMP-GCL algorithm is further extended to 3D scenarios, where user and scatter heights are also estimated. 
    Numerical results validate the theoretical analysis, and verify that MW achieves centimeter- and decimeter-level localization accuracy in 2D and 3D scenarios with only three waveguides.
\end{abstract}

\section{Introduction}

Pinching-antenna systems (PASS) have recently emerged as a promising paradigm for reconfigurable and flexible-antenna architectures \cite{ding2025flexible,liu2025architecture}. 
Originally prototyped by NTT DOCOMO \cite{fukuda2022pinching}, PASS dynamically ``pinch'' small dielectric particles termed pinching antennas (PAs) along a lossless waveguide for signal radiation and reception. 
Compared with conventional fixed-antenna multiple-input multiple-output (MIMO) arrays, PAs' positions can be flexibly tuned along the waveguide over a large spatial range of tens of meters, 
providing new opportunities for near-wired communications.
This nature enables control over large-scale path loss and the spatial radiation pattern, 
enabling adaptive pinching beamforming and blockage-resilient communications \cite{ding2025flexible,liu2025architecture}.

Most existing studies on PASS focused on system modeling and optimization, 
including single-user system sum rate maximization \cite{PASS_RateMaximization}, 
the transmit and pinching beamforming design \cite{wang2025modeling,xu2025joint},  adjustable power radiation \cite{xu2025power}, capacity optimization \cite{ouyang2025capacity}, 
in-waveguide attenuation modeling \cite{xu2025attenuation}, and secure communications \cite{zhu2025secure}. 
However, the channel estimation is challenging since near-field (NF) spherical-wave channels need to be reconstructed for PAs activated at various positions. 
Recently, a deep learning based channel estimation framework for PASS was proposed in  \cite{xiao2025channel}, 
which reconstructs channels from limited pilots using mixture-of-experts and Transformer. 
The authors of \cite{zhou2025mmwave} further proposed an orthogonal matching pursuit (OMP) method using far-end and near-end subarrays for millimeter-wave (mmWave) PASS channel estimation. 
Moreover, a PASS-based indoor positioning was proposed in \cite{zhang2025positioning} using received signal strength indication (RSSI), showing that increasing the number of activated antennas 
enhances accuracy but introduces diminishing returns. 
However, preliminary studies assume a single waveguide, where two subarrays or multiple PAs need to be alternatingly activated to measure channels. 
Furthermore, only colinear PAs can be deployed, which limits the observation range and estimation precision. 

Against this background, this letter proposes a novel user localization and channel estimation framework for PASS.   
The proposed framework groups pre-mounted PAs into multiple subarrays to jointly estimate user and scatterer locations,  
and thus reconstruct PASS channels. 
We consider both single-waveguide (SW) and multi-waveguide (MW) structures. 
Specifically, SW alternatingly activates multiple subarrays along one waveguide, while MW activates one subarray on each waveguide for simultaneous measurement. 
For 2D scenario, we first develop an OMP-based geometry-consistent localization (OMP-GCL) algorithm, which exploits inter-subarray geometric relationships for precise estimation. 
Theoretical analysis on Cram\'{e}r-Rao lower bound (CRLB) reveals that enhancing geometric diversity through mult-subarray deployment improves the estimation accuracy.
Moreover, SW restricts the observing directions within a $180^{\circ}$ half-space range and causes angle ambiguity, while MW resolves this issue and reduces overheads.
We further extend OMP-GCL to the 3D scenario. 
Finally, we provide numerical results to verify the effectiveness. 

\begin{figure}[!t]
    \centering
    \subfloat[SW structure.]{\includegraphics[width=0.4\textwidth]{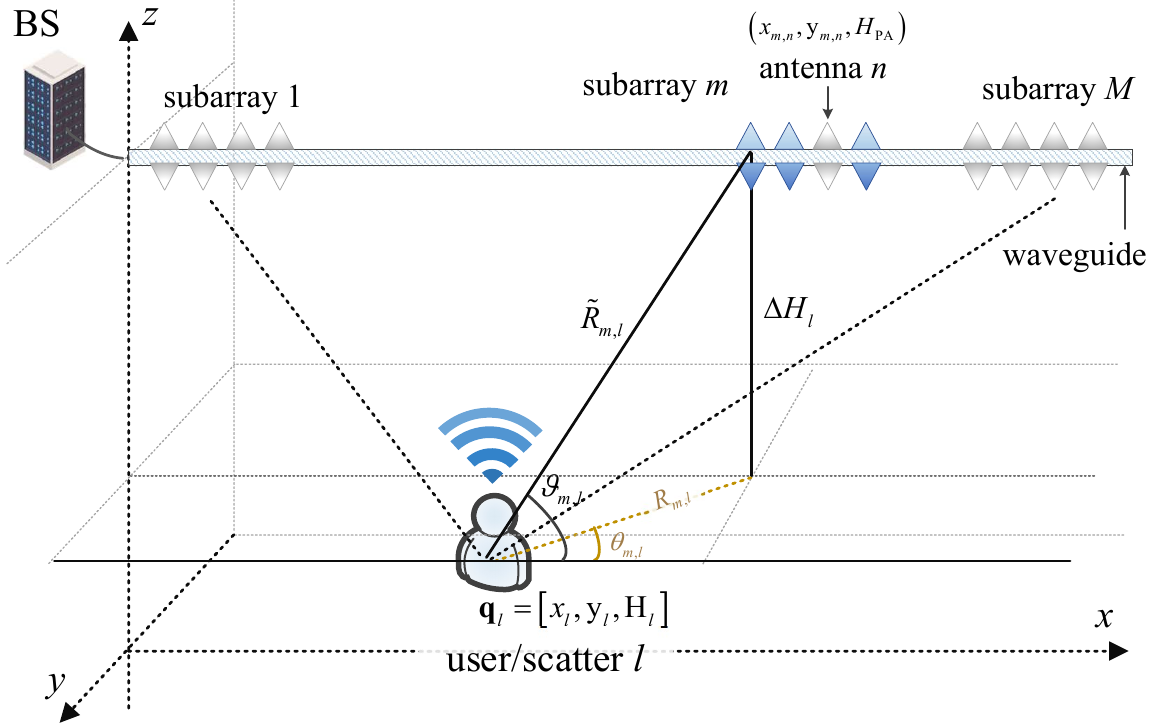}}
    \hfill
    \subfloat[MW structure.]{\includegraphics[width=0.4\textwidth]{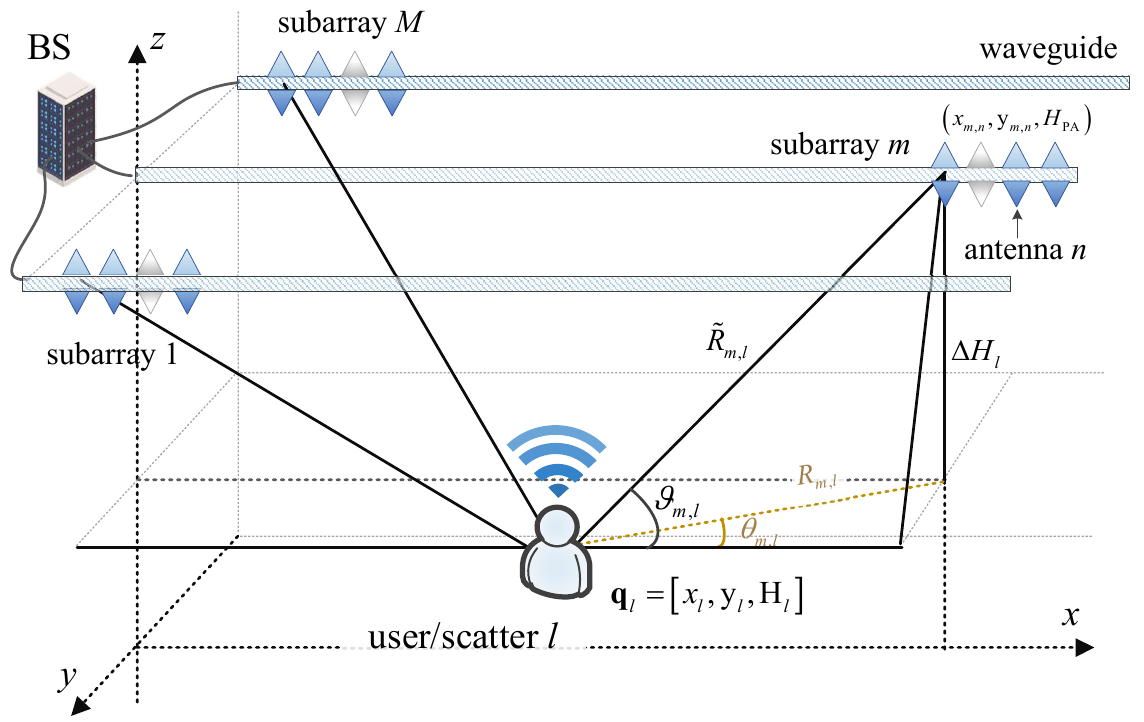}}
    \caption{The proposed joint estimation framework for PASS.}\label{fig_model}
\end{figure}

\section{System Model and Architecture Design}
We investigate a general multi-subarray cooperative estimation framework for PASS, as shown in Fig. \ref{fig_model}. 
$M$ subarrays are equipped by the PASS, and each subarray obtains its individual channel measurements using $N$ PAs. 
The proposed framework adopts either SW or MW structures. 
SW structure deploys all subarrays along a single waveguide that is connected to one RF chain, as shown in Fig. \ref{fig_model}(a). 
MW structure activates the subarrays on different waveguides that are connected to dedicated RF chains, as shown in Fig. \ref{fig_model}(b). 
The discrete activation structure is assumed, where PAs are pre-installed on waveguide(s) for ease of implementation. 
Define $\mathcal{M}$ and $\mathcal{N}$ as the sets of $M$ subarrays and their PAs.
The location of the $n$-th PA of subarray $m$ is $\mathbf{v}_{m,n} = \left(x_{m,n}, y_{m,n}, H_{\mathrm{PA}} \right)$, where $H_{\mathrm{PA}}$ is the fixed height of the waveguide. 
The locations of user/scatters are denoted by $\mathbf{q}_{0} = \left(x_{0}, y_{0}, H_{0}\right)$ 
and $\mathbf{q}_{l} = \left(x_{l}, y_{l}, H_{l}\right)$, $\forall l \geq 1$. 

Each subarray $m$ collects individual measurement through an activation matrix $\mathbf{A}_{m,t}$ 
that randomly activates PAs at time slot $t$. 
Specifically, $\mathbf{A}_{m,t} \!=\! \operatorname{diag}\left(\mathbf{a}_{m,t}\right) 
\!=\! \operatorname{diag}\big(a_{m,0,t}, \allowbreak a_{m,1,t}, \dots, a_{m,N-1,t} \big) \in \mathbb{R}^{N \times N}$ is a diagonal binary matrix, 
where $a_{m,n,t} \in \{0,1\}$ indicates the activation of PA $n$ in subarray $m$ during time slot $t$. 
To obtain individual measurements, 
SW divides the time slots into $M$ subsets as $\mathcal{T} \!=\! \mathcal{T}_{1} \cup \mathcal{T}_{2} \cup \dots \cup \mathcal{T}_{M}$. 
and only activates subarray $m$ during time slots $\mathcal{T}_{m}$, while deactivates the remaining subarrays: 
\begin{equation}
a_{m,n,t}\! \sim\! \mathrm{Bernoulli}(0.5), ~a_{m',n,t} \!=\! 0, ~\forall t \in \mathcal{T}_{m}, ~ \forall m' \!\ne\! m. 
\end{equation}
Moreover, MW enables subarrays to be concurrently activated in each time slot, and obtains individual measurements along each waveguide. 
Hence, the activation for MW is given by $a_{m,n,t}\! \sim\! \mathrm{Bernoulli}(0.5)$, $\forall m \in\mathcal{M}$, $\forall t \in \mathcal{T}$.

\subsection{PASS Signal Model}
The spherical-wave channel for subarray $m$ is modelled by
\begin{equation}\label{channel}
        \mathbf{h}_{m}\!
        \!=\! \mathbf{b}_{m,0}\!\left(\mathbf{q}_{0}\right) + \sum_{l=1}^{L}\mathbf{b}_{m,l}\!\left(\mathbf{q}_{l}\right),     
\end{equation}
including LoS component $\mathbf{b}_{m,0}$ (for $l=0$) and $L$ NLoS components $\mathbf{b}_{m,l}\!\left(\mathbf{q}_{l}\right)$ ($\forall 1\leq l\leq L$). 
Specifically, the LoS/NLoS component $\mathbf{b}_{m,l}\!\left(\mathbf{q}_{l}\right)$ is given by
\begin{multline}\label{response}
       \mathbf{b}_{m,l}\!\left(\mathbf{q}_{l}\right)\!=\!
        \big[\alpha_{m,0,l} e^{-j\kappa r_{m,0,l}},
        \alpha_{m,1,l}{e^{-j\kappa r_{m,1,l}}},\dots, \\
        \alpha_{m,N-1,l}{e^{-j\kappa r_{m,N-1,l}}}\big]^{\top}, \quad
        \forall l \in \{0,1,\dots,L\},
\end{multline}
where $\kappa \!\triangleq\! \tfrac{2\pi}{\lambda}$ is the wave-domain number. 
$r_{m,n,l} \!=\! \sqrt{(x_{m,n} \!-\! x_{l})^2 \!+\! (y_{m,n}-y_{l})^2 \!+\! \Delta H_{l}^2}$ is the distance from PA $n$ at waveguide $m$ to the user/scatter located at $\mathbf{q}_{l}$ 
with $\Delta H \!\triangleq\! H_{\mathrm{PA}}-H_{l}$. 
$\alpha_{m,n,0} \!=\! \tfrac{\lambda}{4\pi r_{m,n,0}}$ is the complex path coefficient of the LoS component, and 
$\alpha_{m,n,l} \!=\! \tfrac{\lambda e^{-j\kappa r_{l}^{\mathrm{su}}}}{(4\pi)^{3/2} r_{m,n,l} r_{l}^{\mathrm{su}}}$, $\forall l\geq 1$, is the complex path coefficient of the $l$-th NLoS component 
with $r_{l}^{\mathrm{su}}$ being the scatter-user distance. 
\begin{remark}
    By estimating the user and scatter locations $\mathbf{Q} \!=\! \big[\mathbf{q}_{0}, \mathbf{q}_{1},\dots, \mathbf{q}_{N}\big]^{\top}$, 
    the spherical-wave channels $\mathbf{h}_{m}$ can be reconstructed for arbitrary PA placements.
\end{remark}

The in-waveguide signal propagation vector $\mathbf{g}_m\!\in\! \mathbb{C}^{N\times 1}$ is 
\begin{equation}
\mathbf{g}_m \!=\! \big[ e^{j\kappa  n_{\mathrm{eff}} x_{m,0}},e^{j\kappa  n_{\mathrm{eff}} x_{m,1}}, \dots, e^{j\kappa  n_{\mathrm{eff}} x_{m,N-1}} \big]^\top,
\end{equation}
where $n_{\mathrm{eff}}$ is the guided refraction index.
At each time slot $t$, the pilot signal received by subarray $m$ is given by
\begin{equation}
y_{m,t}^{\mathrm{pilot}} = \sqrt{P_0} \mathbf{g}_{m}^H \mathbf{A}_{m,t} \mathbf{h}_{m} x_{t}^{\mathrm{pilot}} + n_{m,t},
\end{equation}
where $n_{m,t}$ is the additive white Gaussian noise (AWGN).

\subsection{Sparse Estimation}
Stacking channel measurements over $T$ time slots, the observation at subarray $m$ using $x_{t}^{\mathrm{pilot}}=1$ is given by
\begin{equation}
    \begin{split}
        \mathbf{y}_m^{\mathrm{pilot}} 
        & \!=\! 
        \big[\mathbf{A}_{m,1}\mathbf{g}_{m}, \mathbf{A}_{m,2}\mathbf{g}_{m}, \dots, \mathbf{A}_{m,T}\mathbf{g}_{m}\big]^{\mathsf{H}} \mathbf{h}_{m}\!+\! \mathbf{n}_{m} \\
        & \!=\! \mathbf{W}_m \mathbf{h}_m \!+\! \mathbf{n}_{m},
    \end{split}
\end{equation}
where $\mathbf{y}_{m}^{\mathrm{pilot}} \!=\! [y_{m,1}^{\mathrm{pilot}},y_{m,2}^{\mathrm{pilot}},\dots,y_{m,T}^{\mathrm{pilot}}]^\top$ stacks the observed signals, 
$\mathbf{W}_{m} \!=\! \big[\mathbf{A}_{m,1}\mathbf{g}_{m}, \mathbf{A}_{m,2}\mathbf{g}_{m}, \dots, \mathbf{A}_{m,T}\mathbf{g}_{m}\big]^{\mathsf{H}}\! \in\mathbb{C}^{T\times N}$  is the measurement matrix,
and $\mathbf{n}_m \!=\! [n_{m,1},\dots,n_{m,T}]^\top$ is the measurement noise.

As indicated in \eqref{channel}, in order to reconstruct the spherical-wave channels given the arbitrary PA positions, 
the positions  $\mathbf{Q}=\{\mathbf{q}_{l}\}$ of the user and $L$ scatters 
should be accurately estimated from $\left\{\mathbf{y}_{m}^{\mathrm{pilot}}\right\}$ and $\left\{\mathbf{W}_{m}\right\}$. 
To this end, one can construct a distance-angle dictionary $\mathbf{\Psi}_{m}$, 
and employ sparse recovery algorithms (e.g., OMP) to identify the most likely user positions from the measurement vector $\mathbf{y}_m^{\mathrm{pilot}}$ that fits
\begin{equation}\label{measurement}
\mathbf{y}_m^{\mathrm{pilot}} = \bm{\Psi}_m \mathbf{u}_m + \mathbf{n}_m,
\end{equation}
where $\mathbf{u}_m$ is a sparse coefficient vector with non-zero entries indicating the active atoms corresponding to the estimated user location. 
Based on the compressed sensing (CS) theory, the sparse vector $\mathbf{u}$ is obtained by
\begin{equation}
    \mathbf{u}_m = \arg\min_{\mathbf{u}} \|\mathbf{y}_m^{\mathrm{pilot}} - \bm{\Psi}_m \mathbf{u}\|_2^2, \quad \text{s.t.} \quad \|\mathbf{u}\|_0 \leq 1.
\end{equation}

There are mainly two methods to construct dictionary $\bm{\Psi}_{m}$, 
i.e., the NF polar-domain dictionary \cite{cui2022nearfield} and the distance-parameterized (DP) angle-domain dictionary \cite{zhang2024distanceparameterized}. 
For the polar-domain dictionary, the PAs can be deployed to form large effective aperture and thus achieve near-field propagations. 
Then, the angle-distance parameters can be estimated to reconstruct spherical wavefront for a fixed array. 
However, the polar-domain dictionary suffers from high correlations between the dictionary atoms that represent different user positions \cite{zhang2024distanceparameterized}. 
Hence, although polar-domain dictionary can estimate accurate spherical-wave channel for a fixed MIMO array, it cannot accurately recover user localization and PASS channels. 
To address this issue, we construct a tailored DP angle-domain dictionary $\bm{\Psi}_{m} \left(R_{m,l}\right)$ in this paper, 
which is an angel-domain dictionary parameterized by a fixed distance $R_{m,l}$. 
To further overcome angular ambiguity and to achieve accurate user localization and channel estimation for PASS, 
we propose the multi-subarray estimation solution in the sequel.

\section{PASS Localization and Channel Estimation}
This section performs joint user localization and channel estimation. 
We first propose OMP-GCL in the 2D scenarios, where the direction $\left\{\theta_{m,l}\right\}$ and the 2D distance $\left\{R_{m,l}\right\}$ are estimated given a fixed user/scatter height $H_{l}$. 
Then, we extend OMP-GCL to the 3D scenarios, where $\{H_{l}\}$ is also estimated.

\subsection{2D Localization and Channel Estimation with Fixed $H_{l}$}
The proposed OMP-GCL method leverages CS theory for direction estimation, and relying on the geometrical diversity of multiple subarrays for localization.
\subsubsection{OMP-based Direction Estimation}
Let $R_{m,l}\!=\!\sqrt{(x_{m} \!-\! x_{l})^2 \!+\! (y_{m}\!-\!y_{l})^2}$  denote the estimated 2D distance between the reference antenna (i.e., the $0$-th antenna) of subarray $m$ and the user (or the $l$-th scatter),
and $\theta_{m,l}$ denotes the according angle. 
Thus, the PA-user distance is rewritten as
\begin{equation}\label{parameterized_dist}
    r_{m,n,l}\!\left(R_{m,l}, \theta_{m,l}\right)
    \!=\! \sqrt{R_{m,l}^{2}\!+\!n^{2}d^{2}\!-\!2ndR_{m,l}\cos\theta_{m,l}\!+\!\Delta H_{l}^{2}}.
\end{equation}
By fixing $R_{m,l}$, we construct the DP angle-domain dictionary for each subarray $m$ by
\begin{equation}
    \bm{\Psi}_{m} \!=\! \left\{\bm{\psi}\!\left(R_{m,l},\theta\right)\right\}_{\theta \in \Theta} \!\in\! \mathbb{C}^{N \times G_\theta},
\end{equation}
where $\theta\in\Theta$ is a angular grid point, and $\Theta$ denotes the angular set of $G_\theta$ candidate grid points.
Moreover, $\bm{\psi}\left(R_{m,l},\theta\right) \!=\! \big[{\psi}_{n}\left(R_{m,l},\theta\right)\big]\in\mathbb{C}^{N\times 1}$ 
denotes an atom in the dictionary, where the $n$-th element is given by
\begin{equation}
    \psi_{n}\left(R_{m,l},\theta\right) 
    = \frac{1}{\sqrt{N}}
    \frac{e^{-j\kappa r_{m,n}\left(R_{m,l}, \theta\right)}}{r_{m,n}\left(R_{m,l}, \theta\right)}.
\end{equation}
The measurement-domain atom is obtained by projecting $\bm{\psi}_{g}$ through the measurement matrix $\mathbf{W}_m$ using
\begin{equation}
\bm{\phi}_{m,g} = \mathbf{W}_{m} \bm{\psi}_{g} \in \mathbb{C}^{T \times 1}.
\end{equation}
Then, the overcomplete dictionary $\bm{\Phi}_m \in \mathbb{C}^{T \times G_\theta}$ is formed by concatenating all the $\ell_2$-normalized unit-norm atoms as
\begin{equation}
\bm{\Phi}_m \!=\! \left[ \tfrac{\bm{\phi}_{m,1}}{\|\bm{\phi}_{m,1}\|_2},\ \tfrac{\bm{\phi}_{m,2}}{\|\bm{\phi}_{m,2}\|_2}, \dots, 
\tfrac{\bm{\phi}_{m,G_{\theta}}}{\|\bm{\phi}_{m,G_{\theta}}\|_2} \right].
\end{equation}
OMP iteratively selects the dictionary atom with the highest correlation to the current residual, which is given by
\begin{equation}
    g^\ast = \arg\max_{g} |\bm{\phi}_{m,g}^H \mathbf{y}_{m}^{\mathrm{res}}|,
\end{equation}
where $\mathbf{y}_{m}^{\mathrm{res}} = \mathbf{y}_m^{\mathrm{pilot}} - \sum_{i=0}^{l}\mathbf{b}_{m,i}$ is the residual signal.
The corresponding $\varphi_{m} = \cos\theta_{g^{*}}$ is extracted from the $g^\ast$-th angle on the discretized angular grid. 
The signal coefficient can be  estimated via least squares, i.e., 
$\hat{u}_{m,l} = \bm{\phi}_{m,g^{*}}^\dagger \mathbf{y}_{m}^{\mathrm{res}} 
    = \tfrac{1}{\|\bm{\phi}_{m,g^{*}}\|_2^2}\bm{\phi}_{m,g^{*}}^{\mathsf{H}} \mathbf{y}_m^{\mathrm{res}}$.

\begin{algorithm}[!tp]
    \caption{OMP-GCL Algorithm}
    \label{alg:alt-omp-dcl}
    \begin{algorithmic}[1]
    \Require
    Received signals $\{\mathbf{y}_m\}_{m=1}^M$, 
    sensing matrices $\{\mathbf{W}_m\}$, 
    initial distance $R_m \gets R^{\mathrm{init}}, \forall m$, angle grid $\Theta$.
    \Ensure
    Estimated user position $\mathbf{q}_{0}$, scatter location $\mathbf{q}_{l}$, 
    and channel vector $\mathbf{h}\left(\mathbf{q}\right)$.
    \For{$l=0,1,\dots,L$}
    \Repeat
        \State Construct $\bm{\Phi}_m$ and estimate $\cos\theta_{m,l}$ by OMP, $\forall m$.
        \State Obtain $\mathbf{q}_{l}$ by Algorithm \ref{alg:sign_enum_localization}. Update $R_{m,l}$, $\forall m$.
    \Until{maximum iterations $I$ reached}
    \State Compute $r_{l}^{\mathrm{su}}=\|\mathbf{q}_{l}-\mathbf{q}_{0}\|_{2}$ if $l\geq 1$.
    \State Compute residual $\mathbf{y}_{m}^{\mathrm{res}}=\mathbf{y}_{m}^{\mathrm{pilot}}-\sum_{i=0}^{l}\mathbf{b}_{m,l}(\mathbf{p}_{l}).$
    \EndFor
    \State \Return $\mathbf{Q}$, $\mathbf{h}$. 
    \end{algorithmic}
\end{algorithm}

\begin{algorithm}[!tp]
    \caption{Projection Minimization based Localization}
    \label{alg:sign_enum_localization}
    \begin{algorithmic}[1]
    \Require Subarray positions $\{\mathbf{v}_m = [x_m, y_m]^\top\}_{m=1}^M$,  
    direction cosine estimates $\{\varphi_m = \cos\theta_m\}_{m=1}^M$, 
    regularization $\varepsilon > 0$, and penalty weight $\lambda > 0$.
    \Ensure Estimated user/scatter location $(x, y)$ and full angles $\{\hat{\theta}_m\}$.
    \State Construct $\mathcal{S} \in \{-1, +1\}^{M}$ with $2^{M}$ combinations of $\mathbf{s}$.
    \State Initialize the best cost $J_{\mathrm{opt}} \leftarrow \infty$.
    \For{$\mathbf{s} \in \mathcal{S}$}
        \State Compute projection matrices $\mathbf{P}_m$ by \eqref{Orthogonal_proj}, $\forall m$. 
        \State Obtain position $\mathbf{q}^{\ast}$ by \eqref{closed_form_LS} for the given $\mathbf{s}$. 
        \State Compute $J_\mathrm{LS}(\mathbf{q}^{\ast}, \mathbf{s})$ by \eqref{Loss_projection} and compute $J_\mathrm{P}(x_{l}^{*})$.
        \If{$J_\mathrm{LS}(\mathbf{q}^{\ast}, \mathbf{s}) + \lambda J_\mathrm{P}(x_{l}^{*}) < J_{\mathrm{opt}}$}
            \State Update $\mathbf{q} \leftarrow \mathbf{q}^{\ast}$, $J_{\mathrm{opt}} \leftarrow J_\mathrm{LS}(\mathbf{q}^{\ast}, \mathbf{s}) + \lambda J_\mathrm{P}(x_{l}^{*})$.
        \EndIf
    \EndFor
    \State\Return user/scatter location $\mathbf{q}$.
    \end{algorithmic}
\end{algorithm}

\subsubsection{Geometry Consistent Localization}
The OMP-based direction cosine estimation suffers from the \textit{angle ambiguity} issue.
Define $\left\{\varphi_{m,l} = \cos\theta_{m,l}\right\}$, $m=1,2,\dots,M$, as the direction cosines estimated at $M$ subarrays.   
Since only $\cos\hat{\theta}_{m,l}$ is available, the sign of $\theta_{m,l}$, denoted by $s_{m}=\mathrm{sign}(\theta_{m})=\mathrm{sign}(\sin\theta_{m})\in\{-1,+1\}$, is ambiguous, 
resulting in $2^M$ possible direction combinations.
To address the angle ambiguity, 
we determine the actual $\theta_{m,l}$ by minimizing the sum projection distance, thus ensuring multi-subarray geometric consistency. 

\textbf{Projection Distance Minimization for Position Estimation.} 
    Given a possible sign combination $\mathbf{s} = [s_{1}, s_{2}, \dots, s_{M}]^{\top} \in \big\{-1, +1\big\}^M$, 
    we define the candidate unit direction vector between the user and subarray $m$ as $\mathbf{u}_{m} = [\varphi_{m,l},\ s_{m} \sqrt{1 - \varphi_{m,l}^2}]^\top$.
    Then, the orthogonal projection matrix $\mathbf{u}_{m}$ is defined as
    \begin{equation}\label{Orthogonal_proj}
        \mathbf{P}_{m} = \mathbf{I} - \mathbf{u}_{m} \mathbf{u}_{m}^\top. 
    \end{equation}
    The user/scatter position $\mathbf{q} = [x, y]^\top$ is estimated by minimizing the sum of squared orthogonal projection distances: 
    \begin{equation}\label{Loss_projection}
        \min_{\mathbf{q}} ~ J_{\mathrm{LS}}(\mathbf{q},\mathbf{s}) = \sum_{m=1}^M \left\| \mathbf{P}_m  (\mathbf{q}- \mathbf{v}_m) \right\|_2^2,
    \end{equation} 
    where $\mathbf{v}_{m} \triangleq [x_m, y_m]^\top$ denotes the location of each subarray $m$ in the horizontal plane. 
    This convex quadratic programming admits a closed-form solution:
    \begin{equation}\label{closed_form_LS} 
        \mathbf{q}^{\ast} \!=\! \left( \sum_{m=1}^M \mathbf{P}_m \!+\! \varepsilon \mathbf{I} \!\right)^{-1} \!\!\!\!\left( \sum_{m=1}^M \mathbf{P}_m \mathbf{v}_m \right)\!, 
    \end{equation} 
    where $\varepsilon > 0$ is a regularization parameter.

\textbf{Direction-Consistent Penalty for Angle Disambiguation.} 
    Among all $2^M$ candidate configurations, the sign combination that achieves the minimal total cost is selected: 
    \begin{equation} 
        \mathbf{s}^{\ast} = \arg\min_{\mathbf{s} \in \{-1,+1\}^M} J_\text{LS}(\mathbf{q}^{*},\mathbf{s}) + \lambda J_\text{P}(x^{*}). 
    \end{equation} 
    $J_\text{P}(x^{*}) = \sum_{m=1}^M \left[ \min\big(0,\ (x^{*} - x_m)\varphi_m \big) \right]^2$ 
    is a soft penalty that enforces $\mathrm{sign}(x^{*} - x_m) = \mathrm{sign}(\varphi_m)$, thus discouraging inconsistent direction estimation and enhancing robustness. 
    Since RF chains are expensive and $M$ is limited, we enumerate all the candidate combinations of $\mathbf{s}$ to search for the optimal solution. 
    Using the optimal $\mathbf{s}^{\ast}$ and the according $\mathbf{q}^{\ast}$, the $l$-th path component can be constructed based on \eqref{response}. 

Algorithm \ref{alg:alt-omp-dcl} summarizes the overall OMP-GCL process, and Algorithm \ref{alg:sign_enum_localization} presents the GCL algorithm. 
Assume that the estimation error of unit direction $\mathbf{u}_{m}$ for each subarray can be modelled as 
an independent zero-mean Gaussian noise vector with covariance $\sigma^2 \mathbf{I}$. 
We analyze CRLB as follows. 
\begin{lemma}[\textbf{Localization CRLB}]
The CRLB for any unbiased estimator of position $\mathbf{q}$ is given by
\begin{equation}\label{CRLB}
\mathrm{Cov}(\widehat{\mathbf{q}}) \succeq \sigma^2 R_{m,l}^{2} \left( \sum_{m=1}^M \mathbf{P}_m \right)^{-1}.
\end{equation}
\begin{proof}
From definition, we have $\mathbf{u}_{m}=(\mathbf{q}-\mathbf{v}_{m})/\|\mathbf{q}-\mathbf{v}_{m}\|$. The log-likelihood function is
$\log p(\{\widehat{\mathbf{u}}_m\}|\mathbf{q}) = -\frac{1}{2\sigma^2} \sum_{m=1}^M \| \widehat{\mathbf{u}}_m - \mathbf{u}_m \|^2 + \text{const}$.
The Fisher information matrix (FIM) for $\mathbf{q}$ is given by
\begin{equation}
\mathbf{J}(\mathbf{q}) = \frac{1}{\sigma^2} \sum_{m=1}^M \left( \frac{\partial \mathbf{u}_m}{\partial \mathbf{q}} \right)^\top \left( \frac{\partial \mathbf{u}_m}{\partial \mathbf{q}} \right).
\end{equation}
Since
$\tfrac{\partial \mathbf{u}_m}{\partial \mathbf{q}} \!=\! \tfrac{1}{\|\mathbf{q} \!-\! \mathbf{v}_m\|} 
\left( \mathbf{I} \!-\! \mathbf{u}_m \mathbf{u}_m^\top \right) \!\triangleq\! \tfrac{1}{\|\mathbf{q} \!-\! \mathbf{v}_m\|} \mathbf{P}_m$,
we obtain
\begin{equation}
\mathbf{J}(\mathbf{q}) = \frac{1}{\sigma^2} \sum_{m=1}^M \frac{1}{\|\mathbf{q} - \mathbf{v}_m\|^2} \mathbf{P}_m
= \frac{1}{\sigma^2 R_{m,l}^2} \sum_{m=1}^M \mathbf{P}_m. 
\end{equation}
This leads to the CRLB in \eqref{CRLB} and completes the proof. 
\end{proof}
\end{lemma}

\begin{remark}[\textbf{Geometric diversity}]
    From Lemma~1, the largest CRLB component (i.e., the worst-case localization variance) 
    is inversely proportional to the minimum eigenvalue $\lambda_{\min}\!\left(\sum_{m}\mathbf{P}_m\right)$.
    A small $\lambda_{\min}$ occurs when $\{\mathbf{u}_m\}$ are nearly parallel, indicating a poor \textit{geometric diversity} of available observing directions. 
    To enhance $\lambda_{\min}$ and thus reduce the CRLB, subarrays can be placed at spatially separated positions 
    such as the boundary or corner of the service area, so that $\{\mathbf{u}_m\}$ span the widest possible angular range uniformly.    
\end{remark}

\begin{remark}[\textbf{Angle ambiguity of SW}]
    The collinear subarrays in SW confine $\{\mathbf{u}_m\}$ to a $180^{\circ}$ half-plane, leading to identical observations at mirror positions 
    $\mathbf{q}\!=\!(\hat{x}_l,\hat{y}_l,H_l)$ and $\mathbf{q}'\!=\!(\hat{x}_l,-\hat{y}_l,H_l)$, corresponding to $(R_{l},\theta_{l}, H_{l})$ and $(R_{l},-\theta_{l}, H_{l})$.
    Although SW can achieve a low CRLB, it suffers from angle ambiguity in $\mathrm{sign}(\theta_{m})$, which degrades practical localization accuracy.
    In contrast, MW enables $\{\mathbf{u}_m\}$ to 
    span the full $360^{\circ}$ plane and thus uniquely resolves $\mathbf{q}$.
\end{remark}

\subsection{3D Localization and Channel Estimation with Unknown $H$}
In practical scenarios, the elevation $H_{l}$ is typically unknown. 
Hence, we extend OMP-GCL for 3D scenarios, where position $\mathbf{q}_{l} = [x_{l}, y_{l}, H_{l}]^{\top}$ needs to be jointly estimated. 
For each subarray $m$, we parameterize the DP dictionary by the 3D distance $\widetilde{R}_{m,l} \!=\! \sqrt{(x_{l}-x_{m})^{2} \!+\! (y_{l}-y_{m})^{2} 
\!+\! \Delta H_{l}^{2}}$. 
Then, each subarray estimates $\cos\vartheta_{m,l}\!=\!\frac{x_{l}\!-\!x_{m}}{\widetilde{R}_{m,l}}$ 
from residual measurement $\mathbf{y}_{m}^{\mathrm{res}}$.
Specifically, the PA-user distance can be rewritten as 
$r_{m,n,l}\!\left(\widetilde{R}_{m,l},\vartheta_{m,l}\right)\!=\!\sqrt{\widetilde{R}_{m,l}^{2}\!+\!n^{2}d^{2}\!-\!2nd\widetilde{R}_{m,l}\cos\vartheta_{m,l}}$.
Hence, at a specific angle $\vartheta$, the atom of the DP angle-domain dictionary is given by
\begin{equation}
    \psi_{n}\left(\widetilde{R}_{m,l},\vartheta\right) 
    = \frac{1}{\sqrt{N}}
    \frac{\lambda e^{-j\kappa r_{m,n,l}\left(\widetilde{R}_{m,l}, \vartheta\right)}}{4\pi r_{m,n,l}\left(\widetilde{R}_{m,l}, \vartheta\right)}, ~ \vartheta\in{\Theta}.
\end{equation}

Similar to the 2D scenario, the OMP selects the atom with maximum correlation to the observation, 
thus estimating the direction cosine observed by each subarray. 
Based on the obtained $\{\cos\vartheta_{m,l}\}_{m=1}^M$ and the subarray positions $(x_m, y_m, H_{\mathrm{PA}})$, 
the 3D location $\mathbf{q}_{l}$ is estimated by solving the following geometric consistency least-squares problem: 
\begin{equation}
    \min_{x_{l}, y_{l}, z_{l}\geq 0} \quad \sum_{m=1}^{M} \left[ z_{l} \!+\! (y_{l} \!-\! y_m)^2 \!-\! \delta_{m,l} (x_{l} \!-\! x_m)^2 \right]^2, \label{eq:gcl_3d_obj}
\end{equation}
where $\delta_{m,l} = \frac{1}{\cos^2\vartheta_{m,l}} - 1$ is a fixed constant, 
$z_{l}$ is an auxiliary variable for squared height offset, and the user elevation is reconstructed as $H_{l} = H_{\mathrm{PA}} - \sqrt{z_{l}}$. 

\begin{figure}[!tp]
    \centering
    \includegraphics[width=0.95\linewidth]{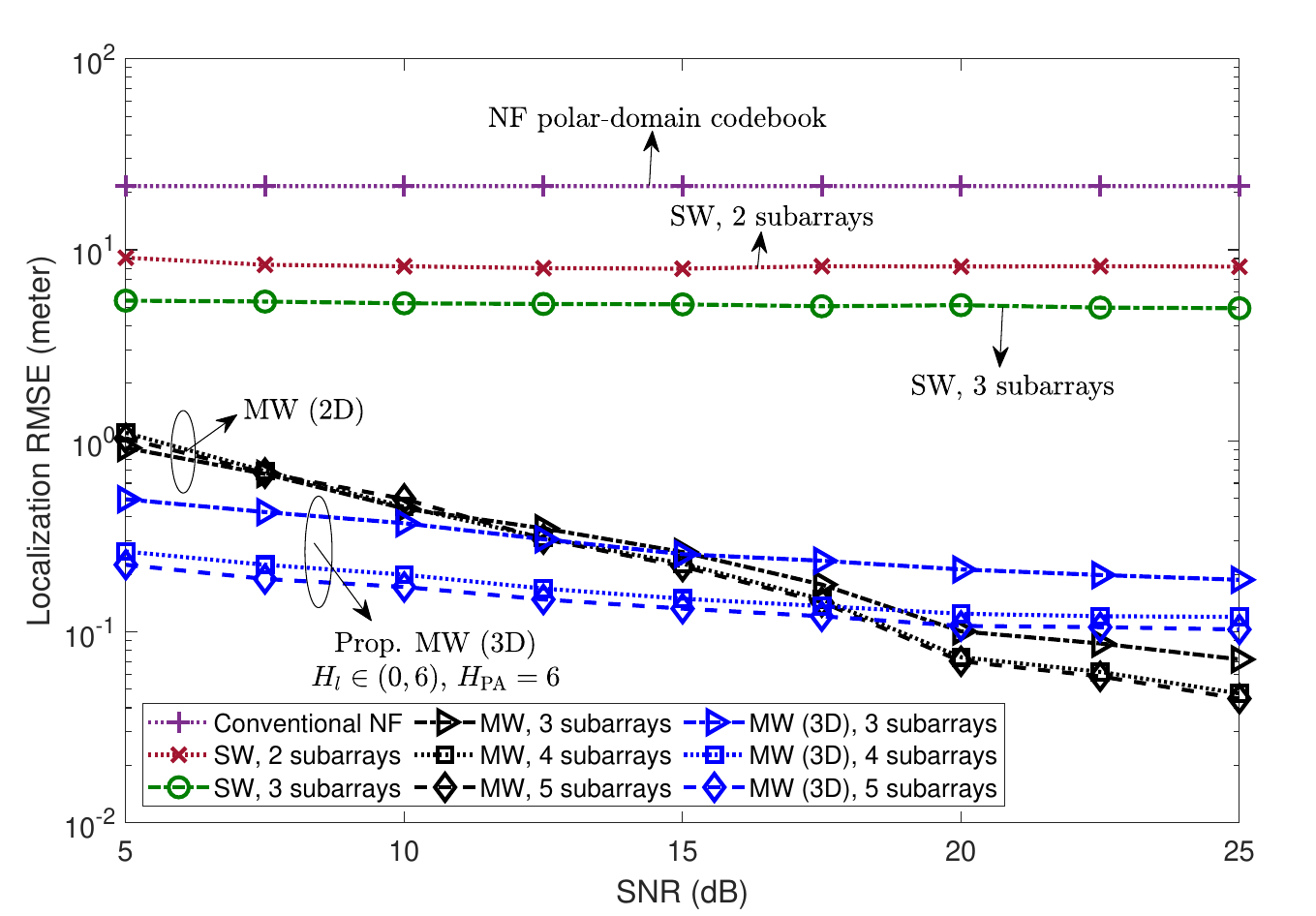}
    \caption{Comparisons of RMSE for user localization.}\label{fig_RMSE}
\end{figure}

\begin{figure}[!tp]
    \centering
    \includegraphics[width=0.95\linewidth]{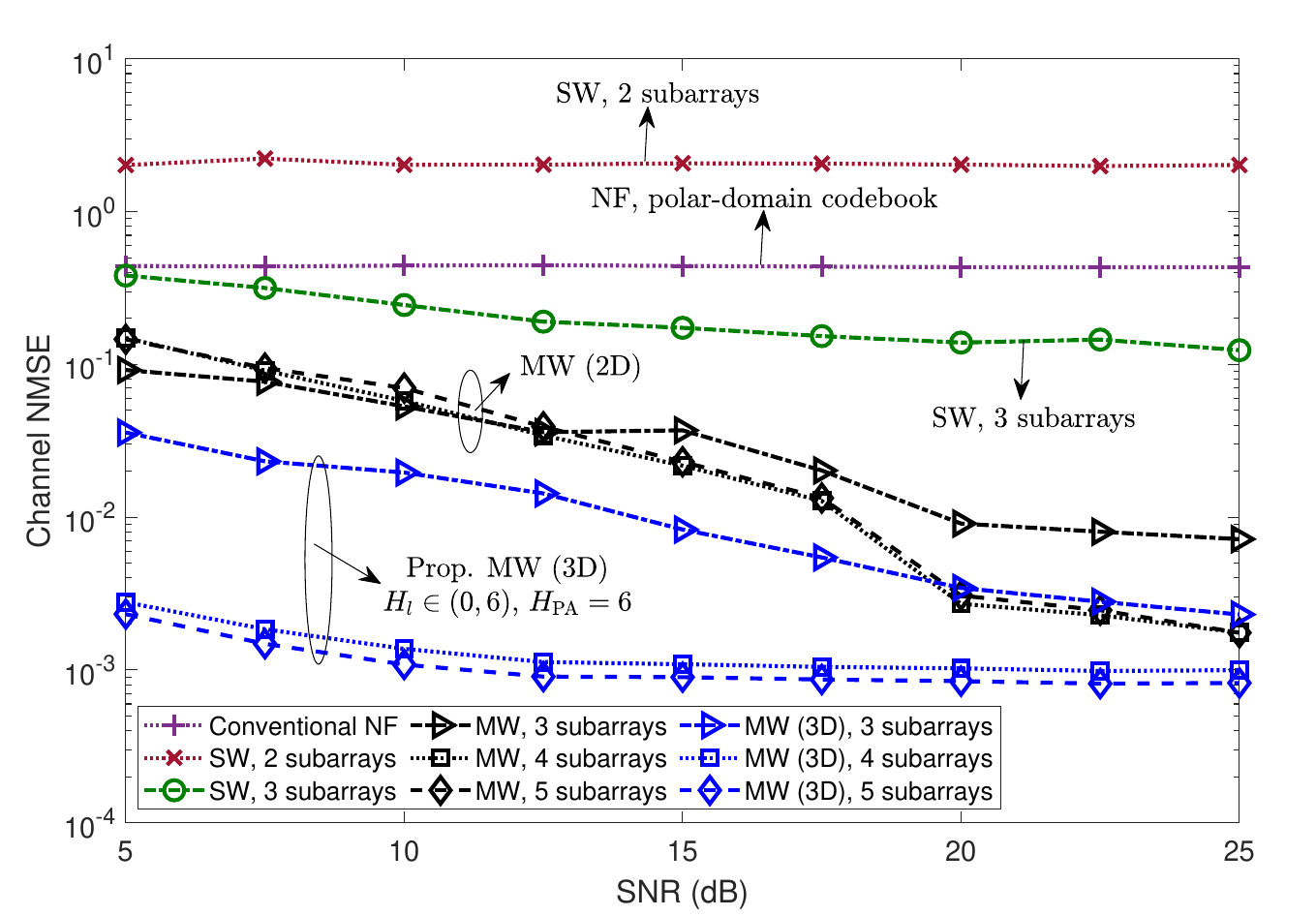}
    \caption{Comparisons of NMSE using different algorithms.} \label{fig_NMSE}
\end{figure} 

\section{Numerical Results}
We provide numerical results in this section. 
The user/scatter is randomly located in an area of $S_{\mathrm{x}}\times S_{\mathrm{y}} ~ \mathrm{m}^{2}$, 
where $S_{\mathrm{x}} = S_{\mathrm{y}} = 30$ m. 
For 2D scenarios, we set $H_{\mathrm{PA}}=2$ m and $H_{l}=0$ m, $\forall l$. 
For 3D scenarios,  we set $H_{\mathrm{PA}}=6$ m and $H_{l}$ is randomly sampled from $H_{l} \in [0, 6]$ m.  
The operating frequency is $f=28$ GHz. 
The PA spacing of each subarray is set as $d=\lambda/2$ m. 
The waveguides are equally spaced with an interval of $D_{\mathrm{y}}=S_{\mathrm{y}}/(M-1)$ m along the $y$-axis. 
Each subarray equips a number of $N=32$ PAs. 
For the SW structure, the $n$-th PA at the $m$-th subarray is located at $x_{m,n} = (m-1)D_{\mathrm{x}}+(n-1)d$ and $y_{m,n} = S_{{\mathrm{y}}}/2$, 
where $D_{\mathrm{x}}=S_{\mathrm{x}}/(M-1)$ is the interval of waveguides over the $x$-axis. 
For the MW structure, PA subarrays are arranged along the boundary of the user region, with the first four subarrays located at the corner points and the others at the edge midpoints.
Two baseline schemes that can be adopted in 2D scenarios are considered. 
(i) \textit{Conventional NF}: This scheme deploys a single subarray with $N=96$ antennas at $(0,S_{{\mathrm{y}}}/2,H_{\mathrm{PA}})$ over one waveguide. 
The polar-domain dictionary \cite{cui2022nearfield} is adopted to estimate user locations and reconstruct spherical-wave channels. 
(ii) \textit{$2$-subarray SW}: Similar to \cite{zhang2024distanceparameterized} and \cite{zhou2025mmwave}, this scheme deploys $M=2$ subarrays along one waveguide and adopts DP dictionary.

Figure \ref{fig_RMSE} presents the root mean square error (RMSE) 
for user localization using different algorithms, where the signal-to-noise ratio (SNR) is set as $\{5, 7.5, 10, \dots, 25\}$ dB. 
As observed, the conventional NF method exhibits a high RMSE, as the high mutual coherence between angular-distance atoms leads to inaccurate localization. 
In comparison, the $2$-subarray SW scheme improves the RMSE as it enjoys low mutual correlation between dictionary atoms, but still suffers from angle ambiguity.  
Benefiting from enhanced spatial diversity and multi-subarray joint estimation, 
the proposed MW scheme achieves the lowest RMSE and significantly outperforms conventional schemes. 
The performance improves as the number of subarrays increases, since the geometric diversity is increased. 
When the SNR is high, the proposed scheme achieves centimeter-level average localization accuracy in 2D scenarios and decimeter-level accuracy in 3D scenarios, 
verifying the potentials of PASS-based localization.

Figure \ref{fig_NMSE} shows the normalized mean-square error (NMSE) of channel estimation versus SNR.
Notably, the SW scheme with two subarrays performs worse than the conventional NF scheme.
This may be because that the angle ambiguity distorts the reconstructed spherical wavefront of SW due to error propagation in distance-angle coupling.
In contrast, although NF polar-domain dictionary leads to inaccurate localization, 
it can still identify a surrogate position with well matched manifold response, resulting in a lower NMSE. 
As the number of subarrays increases, SW significantly improves NMSE due to enhanced spatial diversity.
The proposed MW structure achieves the lowest NMSE by exploiting multi-view geometry and joint estimation among distributed subarrays.



\section{Conclusion}
A joint user localization and channel estimation framework for PASS has been proposed. 
By activating multiple PASS subarrays for channel measurement, 
the locations of user/scatters can be jointly estimated, thus accurately reconstructing channels given arbitrary PA placements. 
Considering 2D scenarios, an OMP-GCL method has been first proposed. 
Each subarray maintained a DP angular-domain dictionary to estimate individual direction cosine, and then 
jointly predicted the locations of user/scatters to overcome angular ambiguity and refine dictionaries. 
The OMP-GCL algorithm was further extended to 3D scenarios.
Numerical results verified that the proposed framework can achieve a high estimation accuracy using only a few waveguides.

\end{document}